# Automated HER2 scoring with uncertainty quantification using lensfree holography and deep learning

Che-Yung Shen[1,2,3†], Xilin Yang[1,2,3†], Yuzhu Li[1,2,3†], Leon Lenk[4], and Aydogan Ozcan[1,2,3*]

[1]Electrical and Computer Engineering Department, University of California, Los Angeles, CA, 90095, USA

[2]Bioengineering Department, University of California, Los Angeles, CA, 90095, USA

[3]California NanoSystems Institute (CNSI), University of California, Los Angeles, CA, 90095, USA

[4]Department of Computer Science, University of California, Los Angeles, CA, 90095, USA

†These authors contributed equally to the work

*Correspondence to: ozcan@ucla.edu

## ABSTRACT

Accurate assessment of human epidermal growth factor receptor 2 (HER2) expression is critical for breast cancer diagnosis, prognosis, and therapy selection; yet, most existing digital HER2 scoring methods rely on bulky and expensive optical systems. Here, we present a compact and cost-effective lensfree holography platform integrated with deep learning for automated HER2 scoring of immunohistochemically stained breast tissue sections. The system captures lensfree diffraction patterns of stained HER2 tissue sections under RGB laser illumination and acquires complex field information over a sample area of ~1,250 mm² at an effective throughput of ~84 mm² per minute. To enhance diagnostic reliability, we incorporated an uncertainty quantification strategy based on Bayesian Monte Carlo dropout, which provides autonomous uncertainty estimates for each prediction and supports reliable, robust HER2 scoring, with an overall correction rate of 30.4%. Using a blinded test set of 412 unique tissue samples, our approach achieved a testing accuracy of 84.9% for 4-class (0, 1+, 2+, 3+) HER2 classification and 94.8% for binary (0/1+ vs. 2+/3+) HER2 scoring with uncertainty quantification. Overall, this lensfree holography approach provides a practical pathway toward portable, high-throughput, and cost-effective HER2 scoring, particularly suited for resource-limited settings, where traditional digital pathology infrastructure is unavailable.

## INTRODUCTION

Digital pathology has substantially advanced histological evaluation, enabling pathologists to manage growing volumes of tissue samples with enhanced precision, reproducibility, and workflow efficiency(*1–3*). Whole-slide imaging systems and digital scanners have facilitated remote assessments, computational analysis pipelines, and streamlined data archival, collectively transforming modern diagnostic practice(*4*). Despite these advantages, the widespread adoption of digital pathology is still partially limited by the cost, speed, and optomechanical complexity of conventional optical systems(*5*). State-of-the-art commercial whole-slide imaging scanners rely on high-numerical-aperture optics, precision motorized stages, and stable illumination and focusing mechanisms, resulting in relatively large system footprints, long acquisition times, and limited accessibility—particularly in resource-limited or decentralized clinical settings. These limitations motivate the development of alternative imaging paradigms that preserve diagnostic relevance while significantly reducing hardware complexity(*6*).

Recent advances in compact optical imaging platforms have opened new opportunities to democratize access to cost-effective histological imaging. Lensfree inline holography, in particular, offers a portable, cost-effective, and high-throughput modality that eliminates the need for objective lenses and mechanical focusing(*7–12*). By directly recording coherent or partially coherent diffraction patterns over large fields of view and digitally recovering sample information, lensfree holographic systems can achieve rapid imaging with minimal mechanical requirements. Prior demonstrations of lensfree holographic systems for biological and label-free imaging further support their potential as practical tools for computational pathology(*13, 14*), especially in environments where infrastructure or cost constraints preclude the use of high-end whole-slide scanners.

Crucially, the viability of these simplified imaging systems has been greatly accelerated by advances in deep learning-based image reconstruction and analysis(*15–27*). Computational imaging approaches combine simplified optical architectures with powerful neural networks capable of compensating for reduced hardware fidelity(*6, 28, 29*). These hybrid systems have demonstrated the ability to restore fine tissue architecture, infer stain-dependent features, and support reliable classification under constrained imaging conditions. Inspired by these hybrid paradigms, this work targets automated scoring of human epidermal growth factor receptor 2 (HER2)—an inherently challenging and clinically significant task that serves as a benchmark for breast cancer (BC) management. Although prior work has shown that automated HER2 scoring in immunohistochemically (IHC) stained BC tissue(*30–32*) can be performed with advanced digital neural networks(*20, 29, 33, 34*), these methods have predominantly relied on high-fidelity whole-slide images (WSIs) acquired via bulky, conventional optical microscopy systems. This highlights a critical opportunity to investigate whether intelligent, data-driven processing can maintain diagnostic reliability when integrated into simplified, lensless imaging architectures. Validating this synergy is important, as it would demonstrate that algorithmic complexity can effectively substitute for high-end optical instrumentation, thereby decoupling diagnostic precision from hardware cost, size and complexity.

Here, we introduce a compact, cost-effective lensfree holographic imaging platform that leverages deep learning to enable rapid, automated HER2 scoring from inline holograms of IHC-stained breast tissue sections. HER2 assessment is a critical component of BC diagnosis and treatment planning, making it a clinically relevant benchmark for evaluating the practical viability of computational optical imaging systems. To address the scalability and accessibility challenges of conventional pathology workflows, our device is designed to capture holographic information of tissue samples over a field of view (FOV) of ~1,250 mm² at an effective throughput of ~84 mm² per minute, enabling high-throughput tissue analysis without bulky optics, mechanical focusing, precision optical alignment or image registration.

We assessed the HER2-scoring performance of the deep-learning–based lensfree holography approach on a blinded test set comprising 412 held-out cores from patients not seen during training. By incorporating a customized ensemble strategy with 5 digital neural networks, we achieved 84.9% testing accuracy across 4 classes (HER2 scores: 0, 1+, 2+, 3+) and 94.8% testing accuracy across 2 classes (HER2 scores: 0/1+, 2+/3+). Furthermore, uncertainty-aware inference using Bayesian Monte Carlo (MC) dropout enabled enhanced prediction, successfully identifying and rejecting ~30.4% of misclassified cores in an autonomous manner, while maintaining a low loss-rate of ~7.2% for correct predictions. Despite operating without lenses, mechanical focusing, high-precision optical alignment, or image registration, our lensfree holographic approach achieves performance comparable to that of conventional brightfield microscopy-based digital pathology systems(*34*).

By integrating a simplified lensfree holography system with advanced neural architectures, this work demonstrates a cost-effective and compact alternative for automated HER2 scoring (see **Table 1**). Beyond establishing a rigorous proof of concept for this specific diagnostic task, our results suggest broader potential for extending cost-effective, lensfree imaging solutions across a wide range of digital pathology applications. Such platforms could serve as cost-effective digital imaging tools for rapid inference of tissue biomarkers, enabling prompt diagnosis of urgent cases or supporting high-volume pathology workflows. Ultimately, this emerging imaging–AI paradigm offers a viable and scalable approach to cancer detection, diagnosis, and treatment, especially in settings where conventional pathology infrastructure and hardware resources are limited.

**RESULTS**

**Lensfree holography setup and HER2 scoring workflow**

**Figure 1** illustrates the overall imaging and processing workflow of the lensfree holography platform. The system consists of (1) a laser source to provide sequential color illumination ($\lambda =$ 630 nm, 530 nm, 450 nm); (2) a test sample slide holder mounted on a customized movable lateral 2D translation stage, (3) a monochrome complementary-metal-oxide-semiconductor (CMOS) image sensor to sequentially capture the images under red, green and blue illuminations, and (4) controlling circuits and software, forming a compact, cost-effective device designed for rapid and large-FOV lensfree holographic imaging; also see **Table 1**. During image acquisition, coherent illumination from the laser source propagates a distance $Z_1$ to the sample plane. The IHC-stained breast tissue section interacts with the incident optical field through cellular and subcellular structures whose scattering and absorption properties reflect underlying HER2 expression. These interactions modulate the transmitted optical field, which subsequently propagates a distance $Z_2$ to the image sensor plane, forming sample-dependent diffraction patterns. The monochrome CMOS image sensor records the resulting interferometric intensity distributions across its full active area. Lateral scanning and image acquisition are controlled via an Arduino-based hardware interface, along with customized control software and a graphical user interface (GUI). The photograph of the experimental setup is shown in **Fig. 1A**, while the captured lensfree hologram of a HER IHC-stained slide and a zoomed-in view of the camera FOV are presented in **Fig. 1C** and **1D**, respectively. A live demonstration of the lensfree scanning holographic imaging and the corresponding GUI for system control is provided in **Video S1** and **Fig. S1**.

**Figure 1b** depicts the digital inference process. Following hologram acquisition, the recorded holograms are digitally transformed into a complex-valued computational representation for neural inference. Specifically, we apply an angular spectrum (AS)–based propagation model(*35*) to numerically transform the encoded optical intensity: the recorded intensity patterns are numerically back-propagated from the sensor plane over a selected axial distance $Z_3$, not necessarily equal to the physical sensor-to-sample distance $Z_2$, to yield a complex-valued field that estimates both amplitude and phase at a plane close to the tissue plane. This transformation does not necessarily reconstruct the actual complex field representation of the tissue sample, due to the lack of a phase retrieval process(*12, 36*); however, it presents a computationally efficient representation of the tissue structural features by producing a six-channel tensor (composed of the real and imaginary parts for each RGB wavelength) which is then fed into a trained classifier based on EfficientNet-B0(*37*) architecture to perform 4-class HER2 score prediction {i.e., 0, 1+, 2+, or 3+}. This approach allows direct, rapid inference from a holographically encoded representation of the tissue sample, without phase-retrieval steps. We also utilized a customized ensemble voting method using five (M=5) neural networks, each trained with different random initializations and data shuffling, to improve the robustness of HER2 scoring. Representative captured holograms and the corresponding classification results are shown in **Fig. 2** and **Supplementary Fig. S2**, demonstrating the effectiveness of automated HER2 scoring directly from lensfree holography. Finally, to enhance diagnostic trustworthiness and reduce the risk of less confident predictions, we employed MC Dropout as a Bayesian approximation technique(*38, 39*) to autonomously quantify prediction uncertainty in our HER2 scoring model and enable selective decisions about which predictions to discard; these results will be detailed and quantified in the following sub-sections.

**Automated HER2 scoring using lensfree holography**

We evaluated the HER2 scoring performance of our deep-learning-based lensfree holography approach using a blinded test set of 412 *unseen* unique patient cores to assess the accuracy of 4-class (HER2 scores: 0, 1+, 2+, 3+) and 2-class (HER2 scores: 0/1+ vs. 2+/3+) classification tasks. As shown in **Fig. 3A**, using a single hologram-based neural network model, we achieved 4-class testing accuracy of 78.2% and 2-class testing accuracy of 88.8%. The 2+ (equivocal) class remained the most challenging, achieving 57.28% (59/103), whereas all other HER2 classes achieved testing accuracies of 80% or higher. This distribution is consistent with known diagnostic difficulties in HER2 scoring; HER2 2+ cores often exhibit subtle or heterogeneous membranous staining, resulting in ambiguity for both human observers and automated algorithms(*40*).

To further improve the prediction robustness and reduce false-negative errors for clinically relevant HER2-positive cores, we implemented a HER2-positive-sensitive ensemble strategy. Specifically, five (M = 5) independently trained network models were utilized, each trained on identical data but with different random initializations and data shuffling to enable model diversity.

During inference, predictions from all models were evaluated using a customized confidence-aware fusion rule rather than simple majority voting on the most confident prediction (see the Materials and Methods section for details). This design reflects the asymmetric clinical cost of classification errors: false-negative HER2 predictions may lead to missed eligibility for HER2-targeted therapy, whereas false-positive predictions primarily result in additional confirmatory testing. Consequently, the proposed fusion strategy prioritizes sensitivity to HER2-positive signals, allowing localized or subtle membranous expression patterns, particularly those that might otherwise be suppressed by majority voting, to be preserved in the final decision.

As shown in **Fig. 3C**, compared with the original single-model results, our customized ensemble voting strategy with M=5 improved the 4-class testing accuracy from 78.2% to 80.8% and the 2-class testing accuracy from 88.8% to 92.5%. The most substantial benefit was observed in the 2+ (equivocal) class, which increased from 57.3% (59/103) to 68.0% (70/103), demonstrating the ensemble's ability to capture subtle membranous staining patterns that are often challenging for both automated systems and human observers. **Figure 3B and D** further show that the area under the receiver operating characteristic curve (AUC) for the 2+ class increased from 0.895 to 0.935, while the AUC for the binary HER2 classification task improved from 0.969 to 0.979, indicating an improved capability to identify HER2-positive tissue cores. This confidence-aware ensemble strategy successfully enhanced diagnostic sensitivity without substantially compromising specificity, particularly for borderline cases and samples with low staining quality.

To benchmark the performance of the lensfree inline holography–based HER2 classification framework against a conventional optical microscopy-based whole slide scanner (AxioScan Z1, Zeiss), we trained and evaluated the classification models using a brightfield microscopy image dataset that matches the spatial resolution used in this study (1024 × 1024 pixels per image patch at 1.67 μm per pixel). Using the same training and digital inference pipeline with the customized ensemble voting (M=5) as in **Fig. 3C,** this brightfield microscopy image baseline achieved 4-class testing accuracy of 82.8% and 2-class testing accuracy of 92.2%, as shown in **Fig. S3**. To further assess inference performance under conditions that preserve the native spatial resolution of the whole slide scanner at sub-micron resolution, we also applied an interleaved subsampling strategy to the original high-resolution brightfield microscopic images acquired with an AxioScan Z1 (Zeiss); see the Materials and Methods section for details. After an additional training process under this higher resolution configuration, also matching the same digital inference pipeline with M=5 ensemble voting strategy, the resulting 4-class testing accuracy slightly improved to ~83.7%, with a 2-class testing accuracy of ~92.7%, as shown in **Fig. S4**. Notably, these brightfield optical microscopy images used for comparison were acquired with commercially available whole-slide scanners that are substantially more expensive and bulkier than our lensfree holography system. Despite this disparity in hardware complexity, size and cost, our lensfree holography-based inference approach achieved a HER-2 classification performance remarkably close to the brightfield microscopy benchmark, while operating without objective lenses, mechanical focusing, or high-precision optical alignment.

**Uncertainty quantification for HER2 scoring**

To further enhance the reliability of the presented approach, we employed MC Dropout, a Bayesian approximation technique(*38*), to autonomously quantify prediction uncertainty in our HER2 scoring model. During inference, dropout layers were kept active with a dropout rate of $R_{drop} = 70\%$, and each test sample was independently evaluated 256 times with MC dropout ($N_{MC}$=256) to generate a distribution of predictions. We then derived a customized figure of merit (FOM) to collectively quantify the uncertainty of the predictions (see the Materials and Methods section for details). FOM captures both the model's confidence and consistency: higher values indicate predictions that are both confident and stable across stochastic perturbations, while lower values indicate predictions that are either uncertain or unstable. As shown in **Fig. 4B**, the diagonal elements of the confusion matrix showed a rather low uncertainty compared to the other elements.

To enable selective predictions with low uncertainty, we used FOM thresholds of [12.1, 16.2, 11.6, 21.2] for HER2 scores [0, 1+, 2+, 3+], which were empirically selected based on the validation set; these FOM thresholds were blindly applied to our test images and were used to reject tissue cores with FOMs below these thresholds. As shown in **Fig. 4C**, this FOM-based autonomous filtering demonstrated effective uncertainty-aware decision-making, improving the 4-class prediction accuracy from 80.8% to 84.9% and the 2-class prediction accuracy from 92.5% to 94.8%. **Figure 4D** shows that the uncertainty quantification achieved an overall correction rate of 30.4% (i.e., more than 30% of all misclassified cores were successfully identified and rejected) while maintaining a low loss-rate of only 7.2% (the portion of the samples that are correctly predicted but discarded). Our uncertainty quantification approach is also effective for critical cases, such as 3+ cores misclassified as 0 (which were 100% rejected by our FOM) and 1+ cores misclassified as 3+ (also 100% rejected). Importantly, this was accomplished with a low loss-rate on correct predictions: only 5.5% of the correctly classified HER2 0 cores, 7.9% of the correctly classified HER2 1+ cores, 7.1% of the correctly classified HER2 2+ cores, and 8.3% of the correctly classified HER2 3+ cores were rejected by our FOM thresholds. To further reduce the loss-rate of correct predictions, we evaluated a second, more conservative set of FOM thresholds ([11.5, 12.0, 10.2, 16.3]). Under this configuration, the loss-rate decreased to 3.6%, while the correction rate was 20.2%, as shown in **Fig. S5**. This trade-off illustrates the flexibility of our uncertainty quantification framework, enabling FOM threshold selection to be tailored to clinical tolerance for rejection versus misclassification. Collectively, these results demonstrate that uncertainty-aware inference enables selective predictions in an interpretable way, allowing the model to abstain from low-confidence decisions and thereby improving the robustness and overall diagnostic reliability of automated HER2 scoring.

**Automated HER2 scoring using monochrome lensfree holography**

Building upon the sequential multi-color illumination approach demonstrated above, we further developed a single-wavelength, monochrome lensfree holographic imaging pipeline that eliminates the need for multi-channel color acquisition, further reducing system cost and increasing imaging throughput. Unlike color holograms, which encode spectral diversity across multiple wavelength bands, monochrome reconstructions rely on a single coherent illumination channel and a cost-effective image sensor. We compared HER2 scoring results obtained from individual color channels with those derived from full RGB holograms, as shown in **Fig. 5A**. Despite the substantial reduction in spectral information and hardware complexity, the monochrome ensemble approach with the blue channel ($\lambda$=450 nm, M=5) achieved a 4-class HER2 testing accuracy of 75.0% and a binary HER2 testing accuracy of 92.5%. **Figure 5B** further shows the corresponding confusion matrix, confirming high sensitivity to HER2-positive cores, with 96 out of 103 HER2 3+ cores correctly identified despite the absence of color information. In addition, under blue light, the AUC for 2-class classification reached 0.978, as shown in **Fig. 5C**.

A notable observation from our monochrome HER2 scoring experiments reported in **Fig. 5** is that, among the three illumination wavelengths evaluated, the blue channel consistently yielded the strongest classification performance. This observation can be attributed to the optical properties of the chromogen used in immunohistochemical staining. Regions containing 3,3'-Diaminobenzidine (DAB) deposits inherently introduce combined amplitude and phase modulation in a holographic imaging system. DAB forms dense, partially opaque organic precipitates with a refractive index slightly higher than that of the surrounding cellular matrix. Under shorter-wavelength blue illumination, these refractive index variations produce a stronger optical signal than at other wavelengths. Consequently, the digitally propagated lensfree holograms exhibit deeper and more sharply defined interference fringes in HER2-positive regions. Importantly, in this regime, the increased absorption of blue light by DAB does not suppress diagnostically relevant information. Instead, the enhanced phase sensitivity at shorter wavelengths amplifies subtle morphological differences at the membrane level, allowing phase-based contrast to dominate even in regions of strong staining. Together, these effects explain why single-wavelength (blue) holography preserves sufficient structural and phase information to support robust HER2 scoring, despite the absence of color information.

**Impact of the digital propagation distance on the automated HER2 scoring performance**

In the analyses presented above, a fixed angular-spectrum (AS) based back-propagation distance ($Z_3$ = 2.4 mm) was used for digital propagation of the lensfree holograms. We further examined how this choice of digital propagation distance ($Z_3$) influences the downstream HER2 classification performance. **Figure 6** shows that classification accuracy exhibited a strong dependence on the selected axial propagation distance, which is closely related to the physical sample-to-sensor distance ($Z_2$ = 2.84 mm). The highest 4-class HER2 testing accuracy was achieved at an axial propagation distance of $Z_3$ = 2.4 mm, reaching 80.8%, followed by 77.8% at $Z_3$ = 2.84 mm. In contrast, shorter propagation distances resulted in reduced classification performance, due to the increased spatial overlap between the real image and the twin image of the tissue structure, reaching accuracies of 75.2% for $Z_3$ = 1.2 mm and 74.8% for $Z_3$ = 0 mm. For the latter, when no digital back-propagation was applied ($Z_3$ = 0 mm), the classification performance for the HER2 2+ class significantly degraded, with the overall accuracy dropping to 56.3%. Collectively, these results highlight the importance of selecting an appropriate axial propagation distance, which is critical for generating complex fields that retain sufficient textural and morphological cues to support robust digital HER2 scoring.

**DISCUSSION**

In this study, we introduced an integrated imaging–AI workflow for automated HER2 scoring, combining lensfree holography with deep learning–based inference. Unlike conventional brightfield microscopes or whole-slide scanners, our cost-effective and compact platform enables rapid lensfree holographic imaging without lenses or mechanical focusing, reducing the optical and mechanical burden that limits throughput and accessibility in standard pathology hardware. Compared with prior automated HER2 scoring studies(*26*, *30*, *33*, *34*), which rely on high-resolution brightfield images, our approach shifts the burden of resolution and contrast recovery from physical optics to computational sensing, enabling robust biomarker classification with compact, cost-effective hardware. We further introduced a HER2-positive-sensitive ensemble strategy, significantly improving accuracy on challenging HER2 2+ cases while avoiding critical mistakes and false positives. Together, these results and analyses establish the first demonstration of HER2 scoring directly from lensfree holography, presenting a robust proof-of-concept for an imaging platform that is faster, smaller, and more affordable than traditional digital pathology systems, while maintaining competitive classification performance with uncertainty quantification.

A potential next step for this platform is the integration of diffractive deep neural networks ($D^2NNs$)(*41*–*46*), which present an intriguing pathway to further accelerate HER2 scoring within lensfree holographic systems. Unlike conventional digital neural networks, which rely on electronic computation following image acquisition, $D^2NNs$ perform all-optical inference through spatially engineered diffractive layers that modulate the optical wavefront as it axially propagates. Prior demonstrations of all-optical diffractive networks have shown their ability to execute complex classification tasks with negligible power consumption(*41*, *47*–*50*). Incorporating such diffractive processors into a lensfree holography pipeline could, in principle, enable tissue classification to occur directly in the optical domain, reducing or eliminating the need for digital image reconstruction, graphics processing unit (GPU)-based inference, or other computational post-processing of image data. This capability would represent a substantial shift toward ultra-fast, low-latency diagnostic assessment, particularly advantageous for high-throughput pathology workflows with limited computational resources. Such optical neural networks with spatially varying point spread functions offer a compelling direction for future systems in which optical imaging, sensing, and inference are co-designed, potentially enabling real-time, hardware-embedded pathology directly at the point of light detection.

## MATERIALS AND METHODS

### Lensfree holography setup

The lensfree holography system was designed as a compact and cost-effective imaging platform optimized for rapid acquisition of inline holograms suitable for downstream deep-learning–based HER2 scoring. The optical configuration comprises a coherent laser source (WhiteLase-Micro; Fianium Ltd, Southampton, UK), a customized 3D-printed slide holder (fabricated using Stratasys, Ltd, Objet30 V3) connected to a 2D lateral positioning stage to enable scanning of sample slides, and a monochrome CMOS image sensor (acA3800-14um, Basler AG, 1.67 μm pixel size, 6.4 mm × 4.6 mm FOV) positioned at a fixed propagation distance ($Z_2$) away from the sample to record inline holograms. The active area of the CMOS image sensor is approximately 0.3 cm²; therefore, mechanical scanning is required to image the full area of a standard pathology slide. To enable large-area imaging, a cost-effective mechanical scanning platform was constructed using a pair of metal screws, linear bearing rods, and linear bearings, with additional 3D-printed components used for structural support, housing, and mechanical joints. Two stepper motors, each with an integrated threaded rod, driven by stepper motor driver chips (AMIS-30543, Pololu), were employed to provide two-dimensional lateral translation of the stage mounted with sample glass slides.

During each test, the sample slide was translated in a raster scanning pattern relative to the fixed illumination and image sensor, acquiring a total of 176 holograms (16 horizontal × 11 vertical positions) with 50% overlap between adjacent frames. This scanning protocol enabled complete coverage of a sample area of ~12.5 cm$^2$ within ~15 min, as demonstrated in **Video S1**. Multi-channel (RGB) holographic measurements were acquired sequentially at three illumination wavelengths (630 nm, 530 nm, and 450 nm). For each wavelength, the entire slide was scanned under monochromatic illumination to acquire a complete hologram dataset. The illumination wavelength was then switched, and the same scanning procedure was repeated. This process was carried out independently for all three wavelengths. The resulting wavelength-specific hologram datasets were subsequently registered and combined before digital propagation to form multi-channel inputs. Notably, sub-pixel–level registration accuracy is *not* required, as the downstream deep-learning models are robust to minor spatial misalignments across the illumination channels, as desired.

An automated control program with a GUI (**Fig. S1**) and a microcontroller (Arduino Micro, Arduino LLC) were developed to manage image acquisition and scanning. The software allows adjustment of camera parameters, communication with the microcontroller to control illumination and sensor power, and precise regulation of the mechanical scanning motion. This integrated control framework synchronizes illumination, sensor readout, and motorized translation, enabling programmable scanning trajectories and fully automated holographic data acquisition with real-time system monitoring. All acquired holograms were stored as raw-intensity images for subsequent digital propagation and classification. A list of system components and their unit costs is provided in **Table 1**. The total cost of the imaging hardware for low-volume acquisition is less than $980, excluding the laser source and the laptop used for overall control and data storage. The tunable laser we used is not included in this cost estimate, as its tunability is not required; it was used solely for a proof-of-concept demonstration. In practical implementations, it will be replaced by three cost-effective laser diodes or even light-emitting diodes (LEDs) for partially coherent RGB illumination. Moreover, the overall hardware cost can be further lowered through higher-volume manufacturing and component integration.

**Data preparation and labeling**

Fifteen unlabeled breast tissue microarray (TMA) slides, each containing ~100–150 cores (~1 mm in diameter), were anonymously obtained from TissueArray(*51*) without any patient identifiers or links, and were IHC-stained for HER2 at the UCLA Translational Pathology Core Laboratory. After staining, the slides were dried and sealed with coverslips prior to imaging. Following image acquisition with the lensfree holographic system, individual tissue cores were localized using a morphology-based image-processing algorithm and cropped from the acquired holographic images. For each tissue core, the captured data were represented as an RGB intensity hologram with 1024 × 1024 pixels. Each extracted region was then matched to its corresponding HER2 score using a pathologist-verified ground-truth spreadsheet, verified by at least 3 certified pathologists, ensuring accurate association between image content and ground-truth HER2 labels. To mitigate inter-slide variability arising from staining heterogeneity and illumination fluctuations, per-channel intensity normalization was applied prior to dataset generation.

The recorded inline holograms were subsequently digitally propagated using an AS–based free-space propagation model(*35*). Specifically, the measured intensity distribution at the sensor plane was digitally back-propagated toward the sample plane by a propagation distance $Z_3$. This digital propagation produced a complex-valued field estimate; there was no phase recovery step implemented to increase the speed of inference. For the network input, we represented the holographically encoded representation as a six-channel tensor by concatenating the real and imaginary components of the complex fields for the R, G, and B illumination wavelengths. This fully automated preprocessing pipeline produced consistently cropped, labeled tissue cores while preserving geometric structure and staining-related features relevant for classification. We discarded tissue cores with poor staining quality (e.g., uneven or weak chromogen deposition, excessive background, or tissue folds) and cores affected by imaging artifacts. This quality-control step ensured that the final dataset contained consistently interpretable cores suitable for reliable training and evaluation. The final dataset consisted of 1,273 images used for training and validation, and 412 images reserved for blind testing evaluation. We enforced a strict patient-level separation between training/validation and testing to assess generalization at the patient level.

Additionally, to benchmark the performance of the proposed lensfree holographic approach against a conventional digital pathology scanner, we used a reference HER2 IHC-stained image dataset composed of brightfield images acquired with a whole-slide scanner (AxioScan Z1, Zeiss) equipped with a ×20/0.8 NA Plan-Apochromat objective(*34*). Each brightfield image in this dataset corresponds directly to a captured hologram from our lensfree holography system, enabling a fair and direct comparison of classification performance with a standard high-resolution slide scanner.

**Classification network architecture and training**

Our HER2 scoring network was built based on the EfficientNet-B0(*37*) structure due to its lightweight architecture and computational efficiency. Derived through neural architecture search (NAS)(*52*), EfficientNet-B0 features depth-wise separable convolutions, multi-kernel mobile inverted bottleneck (MBConv) blocks, squeeze-and-excitation modules, and a hierarchical resolution pyramid. Collectively, these architectural components enable implicit multi-scale feature extraction, allowing the network to capture both high-frequency holographic textures and lower-frequency tissue-level structures. To accommodate the six-channel holographic input, we adapted the first convolutional layer of EfficientNet-B0 from three input channels to six channels. The network was initialized with ImageNet-pretrained weights and fine-tuned using the reconstructed six-channel hologram tensors as input. The network output consisted of class probabilities corresponding to HER2 scores.

To improve the robustness of our predictions, an ensemble of five (M=5) independently trained EfficientNet-B0 models was constructed. Each model $m$ was trained using identical input data but with different random initializations, data shuffling, and augmentation sequences. For a given tissue core, the most confident prediction was selected as the baseline prediction $\hat{y}_{base} = \text{argmax}(s_m)$, where $s_m = \max_c p_c^{(m)}$ denotes the maximum SoftMax confidence score produced by model $m$, and $p_c^{(m)}$ is the predicted confidence for each HER2 class $c$ by the same model $m$. To reduce false-negative errors in clinically relevant HER2-positive cases, an override mechanism was introduced to identify high-confidence predictions of HER2 2+ or 3+ from any member of the M=5 networks. For this aim, we utilized an adaptive confidence threshold, computed as the average of the model-wise maximum confidence $\tau = \frac{1}{M}\sum_{m=1}^{M} s_m$. The final ensemble prediction was then decided by:

$$\hat{y} = \begin{cases} \text{argmax}_{m:\ (\hat{y}_m \in P) \cap (s_m \geq \tau)}(s_m), & if\ \exists m:\ (\hat{y}_m \in P) \cap (s_m \geq \tau) \\ \hat{y}_{base}, & otherwise \end{cases} \quad (1)$$

where $\hat{y}_m$ is the prediction of model $m$, and $P = \{2+, 3+\}$ denote the set of HER2-positive classes. In other words, if any member of the ensemble networks predicts HER2 2+ or 3+ with confidence exceeding the adaptive threshold $\tau$, the most confident prediction from these members is selected as the HER2 score of the tested tissue sample; otherwise, the ensemble defaults to the globally most confident prediction $\hat{y}_{base}$. This confidence-aware ensemble strategy prioritizes sensitivity while maintaining robustness against false low-confidence predictions.

To establish a fair and computationally efficient benchmark for conventional brightfield microscopy-based whole slide imagers, we also trained and evaluated a classification model using an interleaved subsampling–based inference strategy on the brightfield WSI dataset acquired with an AxioScan Z1 (Zeiss). Specifically, each brightfield image (padded to 8192 × 8192 pixels) was partitioned into a grid of non-overlapping 8 × 8-pixel blocks. For each subsampled image, one pixel was selected from each block, resulting in a single 1024 × 1024-pixel image composed of one representative pixel per block. By repeating this process for all possible pixel offsets within the 8 × 8 grid, a total of 64 interleaved subsampled images were generated, each corresponding to a distinct sampling pattern. These 64 subsampled images collectively retain the full spatial information of the original high-resolution WSI data, while enabling efficient batch-based inference by each HER2 network. The subsampled images were processed as a batch of 64 inputs, and their predictions were averaged to yield a final HER2 score for each tissue sample.

Training of all models presented in this work was performed using categorical cross-entropy loss and the AdamW(*53*) optimizer with cosine-annealed learning rate scheduling. Networks were trained with a batch size of 16 for up to 100 epochs, using an initial learning rate of 0.001 and a weight decay of 0.0001. Standard data augmentation techniques, including flipping and rotations, were applied during training to improve generalization. Model selection was based on validation-set balanced accuracy, and early stopping was employed with a patience of 10 epochs to prevent overfitting. All training and inference experiments were conducted on a desktop workstation equipped with a NVIDIA GeForce RTX 3090 GPU, 64 GB of random-access memory, and a 13th Gen Intel Core™ i7 processing unit. Typical training time needed for a model to converge is ~4 hours. The networks were implemented in PyTorch, and ensemble voting inference (M=5) on a single core tissue sample required ~0.4 s.

## Uncertainty quantification and filtering using MC dropout

Prediction uncertainty was quantified using MC dropout as a Bayesian approximation technique(*38*, *39*). During inference, dropout layers were retained with a dropout probability of 70% (uniform for each neuron), and each test sample was randomly evaluated $N_{MC}$=256 times with different dropouts to generate a distribution of stochastic predictions. For each tissue core sample, a FOM for uncertainty quantification was computed by:

$$\text{FOM} = \frac{S_{\text{baseline}}}{\sigma_S} \tag{2}$$

where $S_{\text{baseline}}$ denotes the baseline confidence score (obtained from the corresponding confidence of the prediction with dropout disabled) and $\sigma_S$ represents the standard deviation of the confidence distribution from the stochastic predictions ($N_{MC}$=256). For entries in the confusion matrix with no predictions, the corresponding value is left blank. Class-specific thresholds [12.1, 16.2, 11.6, 21.2] corresponding to HER2 scores [0, 1+, 2+, 3+], respectively, were applied to the uncertainty FOMs. These thresholds were determined on the validation set to achieve >30% correction rate (the fraction of misclassified predictions that were identified), while maximizing the coverage of the retained correct predictions. Predictions with FOMs below the corresponding class threshold were rejected, enabling selective prediction and uncertainty-aware decision making for HER2 scores.

**Supplementary Materials:** This file contains **Figs. S1-S5**.

## REFERENCES

1. P. W. Hamilton, P. Bankhead, Y. Wang, R. Hutchinson, D. Kieran, D. G. McArt, J. James, M. Salto-Tellez, Digital pathology and image analysis in tissue biomarker research. *Methods* **70**, 59–73 (2014).

2. S. W. Jahn, M. Plass, F. Moinfar, Digital Pathology: Advantages, Limitations and Emerging Perspectives. *J. Clin. Med.* **9**, 3697 (2020).

3. S. Rajaganesan, R. Kumar, V. Rao, T. Pai, N. Mittal, A. Sahay, S. Menon, S. Desai, Comparative Assessment of Digital Pathology Systems for Primary Diagnosis. *J. Pathol. Inform.* **12**, 25 (2021).

4. N. Farahani, A. V. Parwani, L. Pantanowitz, Whole slide imaging in pathology: advantages, limitations, and emerging perspectives. *Pathol. Lab. Med. Int.* **7**, 23–33 (2015).

5. O. Kusta, M. Bearman, R. Gorur, T. Risør, J. B. Brodersen, K. Hoeyer, Speed, accuracy, and efficiency: The promises and practices of digitization in pathology. *Soc. Sci. Med.* **345**, 116650 (2024).

6. M. J. Fanous, P. Casteleiro Costa, Ç. Işıl, L. Huang, A. Ozcan, Neural network-based processing and reconstruction of compromised biophotonic image data. *Light Sci. Appl.* **13**, 231 (2024).

7. W. Bishara, H. Zhu, A. Ozcan, Holographic opto-fluidic microscopy. *Opt. Express* **18**, 27499–27510 (2010).

8. D. Tseng, O. Mudanyali, C. Oztoprak, S. O. Isikman, I. Sencan, O. Yaglidere, A. Ozcan, Lensfree microscopy on a cellphone. *Lab. Chip* **10**, 1787–1792 (2010).

9. O. Mudanyali, D. Tseng, C. Oh, S. O. Isikman, I. Sencan, W. Bishara, C. Oztoprak, S. Seo, B. Khademhosseini, A. Ozcan, Compact, lightweight and cost-effective microscope based on lensless incoherent holography for telemedicine applications. *Lab. Chip* **10**, 1417–1428 (2010).

10. A. Greenbaum, W. Luo, T.-W. Su, Z. Göröcs, L. Xue, S. O. Isikman, A. F. Coskun, O. Mudanyali, A. Ozcan, Imaging without lenses: achievements and remaining challenges of wide-field on-chip microscopy. *Nat. Methods* **9**, 889–895 (2012).

11. A. Greenbaum, Y. Zhang, A. Feizi, P.-L. Chung, W. Luo, S. R. Kandukuri, A. Ozcan, Wide-field computational imaging of pathology slides using lens-free on-chip microscopy. *Sci. Transl. Med.* **6**, 267ra175-267ra175 (2014).

12. A. Ozcan, E. McLeod, Lensless Imaging and Sensing. *Annu. Rev. Biomed. Eng.* **18**, 77–102 (2016).

13. T. Liu, Y. Li, H. C. Koydemir, Y. Zhang, E. Yang, M. Eryilmaz, H. Wang, J. Li, B. Bai, G. Ma, A. Ozcan, Rapid and stain-free quantification of viral plaque via lens-free holography and deep learning. *Nat. Biomed. Eng.* **7**, 1040–1052 (2023).

14. Y. Wu, A. Ozcan, Lensless digital holographic microscopy and its applications in biomedicine and environmental monitoring. *Methods* **136**, 4–16 (2018).

15. A. Madabhushi, G. Lee, Image analysis and machine learning in digital pathology: Challenges and opportunities. *Med. Image Anal.* **33**, 170–175 (2016).

16. Y. Rivenson, Z. Göröcs, H. Günaydin, Y. Zhang, H. Wang, A. Ozcan, Deep learning microscopy. *Optica* **4**, 1437–1443 (2017).

17. M. K. K. Niazi, A. V. Parwani, M. N. Gurcan, Digital pathology and artificial intelligence. *Lancet Oncol.* **20**, e253–e261 (2019).

18. Y. Rivenson, T. Liu, Z. Wei, Y. Zhang, K. de Haan, A. Ozcan, PhaseStain: the digital staining of label-free quantitative phase microscopy images using deep learning. *Light Sci. Appl.* **8**, 23 (2019).

19. Y. Rivenson, A. Ozcan, Deep learning accelerates whole slide imaging for next-generation digital pathology applications. *Light Sci. Appl.* **11**, 300 (2022).

20. B. Bai, H. Wang, Y. Li, K. de Haan, F. Colonnese, Y. Wan, J. Zuo, N. B. Doan, X. Zhang, Y. Zhang, J. Li, X. Yang, W. Dong, M. A. Darrow, E. Kamangar, H. S. Lee, Y. Rivenson, A. Ozcan, Label-Free Virtual HER2 Immunohistochemical Staining of Breast Tissue using Deep Learning. *BME Front.* **2022**, 9786242 (2022).

21. X. Wang, S. Yang, J. Zhang, M. Wang, J. Zhang, W. Yang, J. Huang, X. Han, Transformer-based unsupervised contrastive learning for histopathological image classification. *Med. Image Anal.* **81**, 102559 (2022).

22. H. Chen, L. Huang, T. Liu, A. Ozcan, Fourier Imager Network (FIN): A deep neural network for hologram reconstruction with superior external generalization. *Light Sci. Appl.* **11**, 254 (2022).

23. B. Bai, X. Yang, Y. Li, Y. Zhang, N. Pillar, A. Ozcan, Deep learning-enabled virtual histological staining of biological samples. *Light Sci. Appl.* **12**, 57 (2023).

24. H. Chen, L. Huang, T. Liu, A. Ozcan, eFIN: Enhanced Fourier Imager Network for Generalizable Autofocusing and Pixel Super-Resolution in Holographic Imaging. *IEEE J. Sel. Top. Quantum Electron.* **29**, 1–10 (2023).

25. R. J. Chen, T. Ding, M. Y. Lu, D. F. K. Williamson, G. Jaume, A. H. Song, B. Chen, A. Zhang, D. Shao, M. Shaban, M. Williams, L. Oldenburg, L. L. Weishaupt, J. J. Wang, A. Vaidya, L. P. Le, G. Gerber, S. Sahai, W. Williams, F. Mahmood, Towards a general-purpose foundation model for computational pathology. *Nat. Med.* **30**, 850–862 (2024).

26. E. Vorontsov, A. Bozkurt, A. Casson, G. Shaikovski, M. Zelechowski, K. Severson, E. Zimmermann, J. Hall, N. Tenenholtz, N. Fusi, E. Yang, P. Mathieu, A. van Eck, D. Lee, J. Viret, E. Robert, Y. K. Wang, J. D. Kunz, M. C. H. Lee, J. H. Bernhard, R. A. Godrich, G. Oakley, E. Millar, M. Hanna, H. Wen, J. A. Retamero, W. A. Moye, R. Yousfi, C. Kanan, D. S. Klimstra, B. Rothrock, S. Liu, T. J. Fuchs, A foundation model for clinical-grade computational pathology and rare cancers detection. *Nat. Med.* **30**, 2924–2935 (2024).

27. Y. Zhang, L. Huang, N. Pillar, Y. Li, H. Chen, A. Ozcan, Pixel super-resolved virtual staining of label-free tissue using diffusion models. *Nat. Commun.* **16**, 5016 (2025).

28. D. Choudhury, J. M. Dolezal, E. Dyer, S. Kochanny, S. Ramesh, F. M. Howard, J. R. Margalus, A. Schroeder, J. Schulte, M. C. Garassino, J. N. Kather, A. T. Pearson, Developing a low-cost, open-source, locally manufactured workstation and computational pipeline for automated histopathology evaluation using deep learning. *eBioMedicine* **107** (2024).

29. M. J. Fanous, C. M. Seybold, H. Chen, N. Pillar, A. Ozcan, BlurryScope enables compact, cost-effective scanning microscopy for HER2 scoring using deep learning on blurry images. *Npj Digit. Med.* **8**, 506 (2025).

30. S. Loibl, L. Gianni, HER2-positive breast cancer. *The Lancet* **389**, 2415–2429 (2017).

31. M. Zubair, S. Wang, N. Ali, Advanced Approaches to Breast Cancer Classification and Diagnosis. *Front. Pharmacol.* **11** (2021).

32. B. Smolarz, A. Z. Nowak, H. Romanowicz, Breast Cancer—Epidemiology, Classification, Pathogenesis and Treatment (Review of Literature). *Cancers* **14**, 2569 (2022).

33. S. Wu, M. Yue, J. Zhang, X. Li, Z. Li, H. Zhang, X. Wang, X. Han, L. Cai, J. Shang, Z. Jia, X. Wang, J. Li, Y. Liu, The Role of Artificial Intelligence in Accurate Interpretation of HER2 Immunohistochemical Scores 0 and 1+ in Breast Cancer. *Mod. Pathol.* **36**, 100054 (2023).

34. S. Y. Selcuk, X. Yang, B. Bai, Y. Zhang, Y. Li, M. Aydin, A. F. Unal, A. Gomatam, Z. Guo, D. M. Angus, G. Kolodney, K. Atlan, T. K. Haran, N. Pillar, A. Ozcan, Automated HER2 Scoring in Breast Cancer Images Using Deep Learning and Pyramid Sampling. *BME Front.* **5**, 0048 (2024).

35. K. Matsushima, T. Shimobaba, Band-Limited Angular Spectrum Method for Numerical Simulation of Free-Space Propagation in Far and Near Fields. *Opt. Express* **17**, 19662–19673 (2009).

36. O. Mudanyali, D. Tseng, C. Oh, S. O. Isikman, I. Sencan, W. Bishara, C. Oztoprak, S. Seo, B. Khademhosseini, A. Ozcan, Compact, lightweight and cost-effective microscope based on lensless incoherent holography for telemedicine applications. *Lab. Chip* **10**, 1417–1428 (2010).

37. M. Tan, Q. Le, "Efficientnet: Rethinking model scaling for convolutional neural networks" in *International Conference on Machine Learning* (PMLR, 2019; https://proceedings.mlr.press/v97/tan19a.html?ref=ji), pp. 6105–6114.

38. Y. Gal, Z. Ghahramani, Dropout as a Bayesian Approximation: Representing Model Uncertainty in Deep Learning. arXiv arXiv:1506.02142 [Preprint] (2016). https://doi.org/10.48550/arXiv.1506.02142.

39. A. Goncharov, R. Ghosh, H.-A. Joung, D. D. Carlo, A. Ozcan, Autonomous Uncertainty Quantification for Computational Point-of-care Sensors. arXiv arXiv:2512.21335 [Preprint] (2025). https://doi.org/10.48550/arXiv.2512.21335.

40. T. A. Thomson, M. M. Hayes, J. J. Spinelli, E. Hilland, C. Sawrenko, D. Phillips, B. Dupuis, R. L. Parker, HER-2/neu in breast cancer: interobserver variability and performance of immunohistochemistry with 4 antibodies compared with fluorescent in situ hybridization. *Mod. Pathol. Off. J. U. S. Can. Acad. Pathol. Inc* **14**, 1079–1086 (2001).

41. X. Lin, Y. Rivenson, N. T. Yardimci, M. Veli, Y. Luo, M. Jarrahi, A. Ozcan, All-optical machine learning using diffractive deep neural networks. *Science* **361**, 1004–1008 (2018).

42. Y. Luo, D. Mengu, N. T. Yardimci, Y. Rivenson, M. Veli, M. Jarrahi, A. Ozcan, Design of task-specific optical systems using broadband diffractive neural networks. *Light Sci. Appl.* **8**, 112 (2019).

43. M. Veli, D. Mengu, N. T. Yardimci, Y. Luo, J. Li, Y. Rivenson, M. Jarrahi, A. Ozcan, Terahertz pulse shaping using diffractive surfaces. *Nat. Commun.* **12**, 37 (2021).

44. Ç. Işıl, D. Mengu, Y. Zhao, A. Tabassum, J. Li, Y. Luo, M. Jarrahi, A. Ozcan, Super-resolution image display using diffractive decoders. *Sci. Adv.* **8**, eadd3433 (2022).

45. J. Li, T. Gan, Y. Zhao, B. Bai, C.-Y. Shen, S. Sun, M. Jarrahi, A. Ozcan, Unidirectional imaging using deep learning–designed materials. *Sci. Adv.* **9**, eadg1505 (2023).

46. C.-Y. Shen, J. Li, T. Gan, Y. Li, M. Jarrahi, A. Ozcan, All-optical phase conjugation using diffractive wavefront processing. *Nat. Commun.* **15**, 4989 (2024).

47. J. Li, D. Mengu, Y. Luo, Y. Rivenson, A. Ozcan, Class-specific differential detection in diffractive optical neural networks improves inference accuracy. *Adv. Photonics* **1**, 046001 (2019).

48. J. Li, D. Mengu, N. T. Yardimci, Y. Luo, X. Li, M. Veli, Y. Rivenson, M. Jarrahi, A. Ozcan, Spectrally encoded single-pixel machine vision using diffractive networks. *Sci. Adv.* **7**, eabd7690 (2021).

49. M. S. S. Rahman, J. Li, D. Mengu, Y. Rivenson, A. Ozcan, Ensemble learning of diffractive optical networks. *Light Sci. Appl.* **10**, 14 (2021).

50. B. Bai, Y. Li, Y. Luo, X. Li, E. Çetintaş, M. Jarrahi, A. Ozcan, All-optical image classification through unknown random diffusers using a single-pixel diffractive network. *Light Sci. Appl.* **12**, 69 (2023).

51. Human Paraffin Embedded Tissue Array, *TissueArray.Com*. https://www.tissuearray.com/tissue-arrays.

52. B. Zoph, Q. V. Le, Neural Architecture Search with Reinforcement Learning. arXiv arXiv:1611.01578 [Preprint] (2017). https://doi.org/10.48550/arXiv.1611.01578.

53. I. Loshchilov, F. Hutter, Decoupled Weight Decay Regularization. arXiv arXiv:1711.05101 [Preprint] (2019). https://doi.org/10.48550/arXiv.1711.05101.

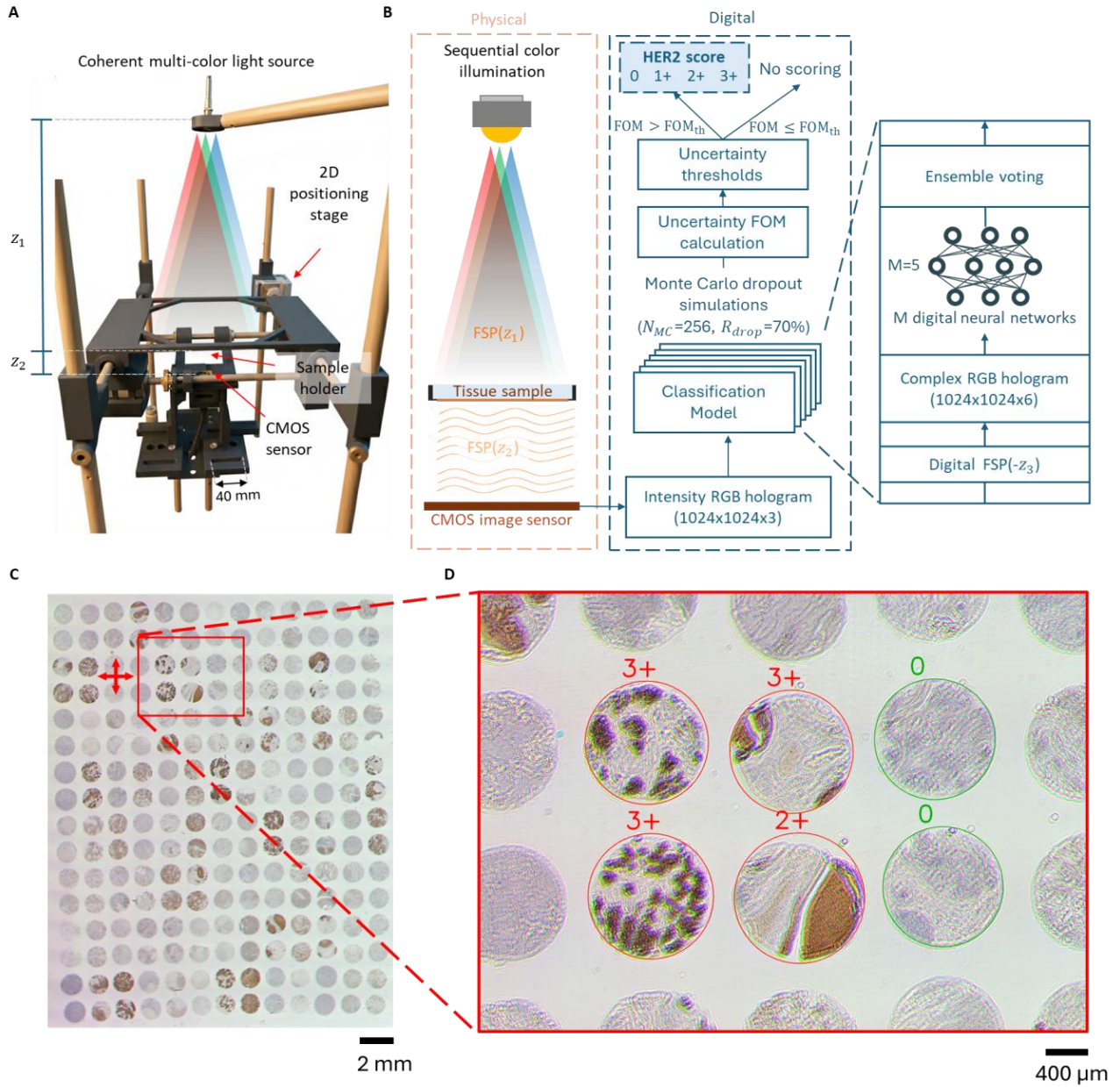

**Fig. 1. Schematics of the customized lensfree holography system and the digital HER2 classification model.** (A) Photo of the lensfree holography system. (B) Physical and digital workflow of the automated HER2 scoring system using lensfree holography. (C) Visualization of the whole-slide lensfree holographic image from a HER2 IHC-stained sample slide. (D) Zoomed-in FOV of the lensfree hologram. Sub-pixel–level spatial registration between the illumination color channels is not required, as the downstream deep-learning models are designed to be robust to spatial misalignments across the RGB illumination channels; hence, rainbow artifacts are observed in each tissue core.

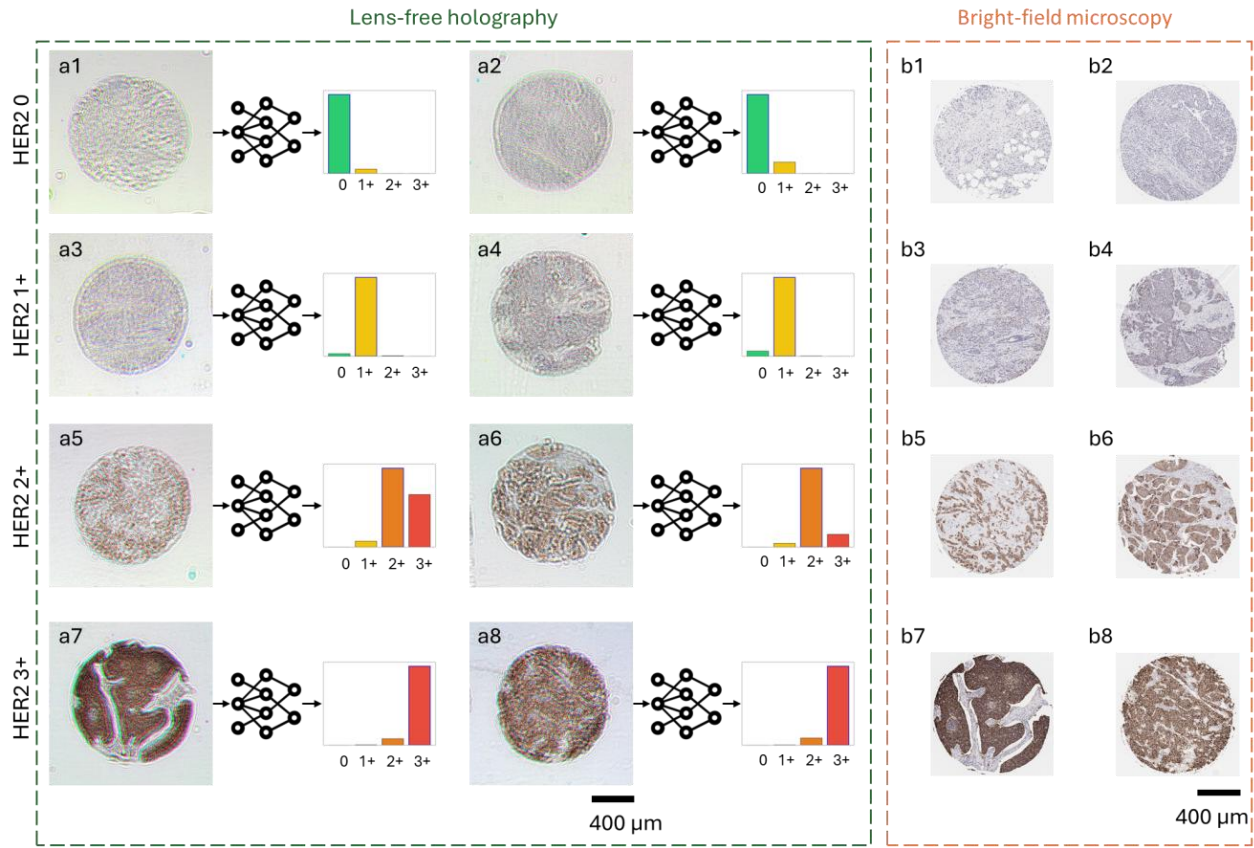

**Fig. 2. Exemplified images of the lensfree holograms and the corresponding brightfield microscopy images.** a1-a8 are the color holograms captured by our lensfree holography setup, each with its network prediction (HER2 score), and b1-b8 are the corresponding brightfield microscopic images for the same patient tissue samples (used only for comparison).

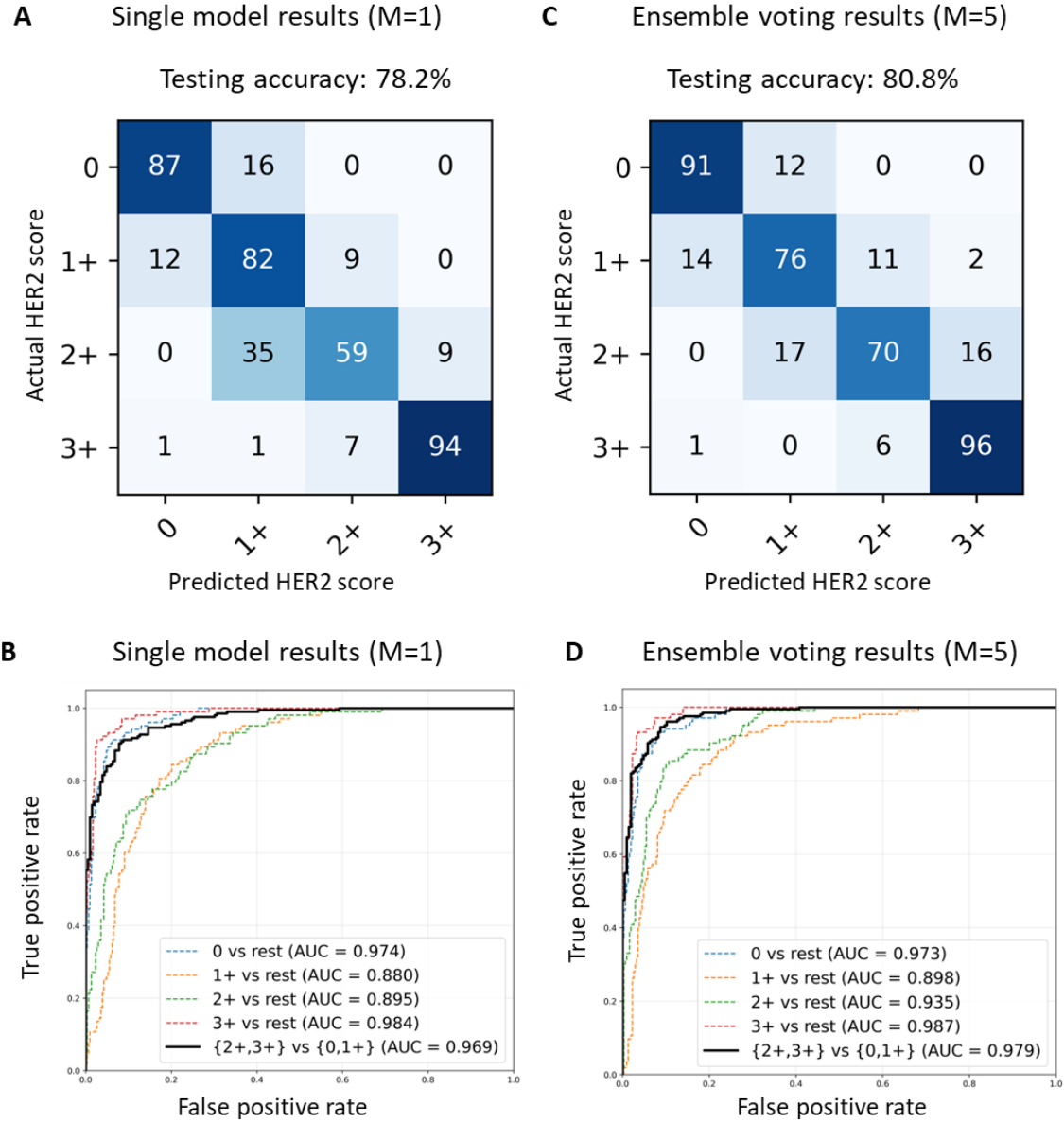

**Fig. 3. Experimental results of the digital HER2 scoring using lensfree holography.** (A) Confusion matrix for the 4-class HER2 testing results using a single digital neural network. (B) ROC curves of the blind testing results using a single digital neural network. (C) Confusion matrix for the 4-class HER2 testing results using the ensemble voting approach with M=5 digital neural networks. (D) ROC curves of the blind testing results using the ensemble voting approach with M=5 digital neural networks.

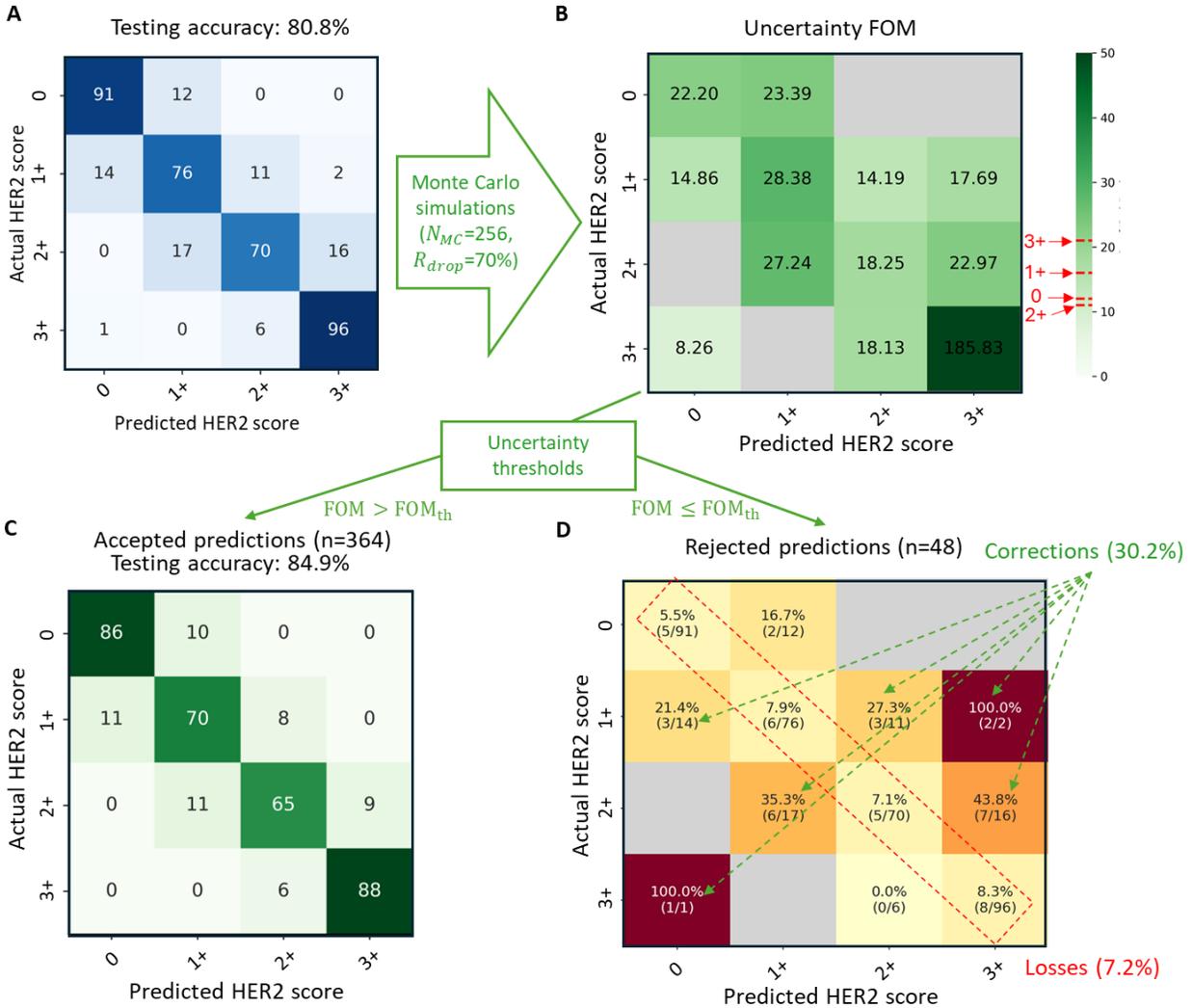

**Fig. 4. Uncertainty quantification of the digital HER2 scoring using MC dropout.** (A) The original 4-class confusion matrix of the ensemble voting results from **Fig. 3c**. (B) The average uncertainty FOM for each element of the classification confusion matrix after the MC dropout-based uncertainty quantification ($N_{MC}$=256, $R_{drop}$=70%). Class-specific FOM thresholds are indicated by red dashed lines on the color bar. (C) Accepted HER2 predictions based on the uncertainty quantification, i.e., FOM > $FOM_{th}$. (D) Rejected HER2 predictions based on the uncertainty quantification, i.e., FOM ≤ $FOM_{th}$. For entries in the confusion matrix with no predictions, the corresponding value is left blank (shown in gray).

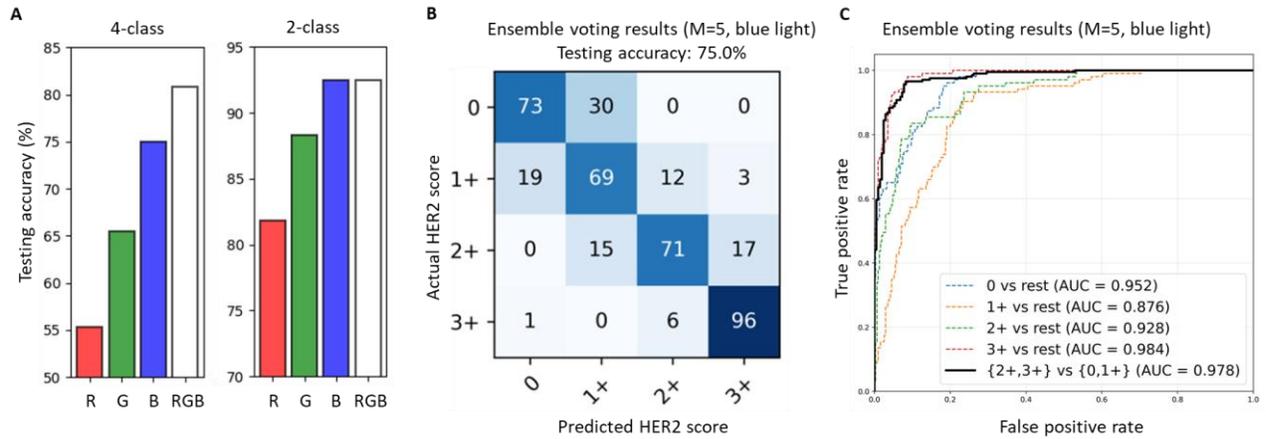

**Fig. 5. Experimental results of the digital HER2 scoring using monochrome lensfree holography.** (A) 4-class and 2-class HER2 testing accuracies using monochrome holograms from three different color channels, along with the RGB baseline. (B) Confusion matrix of the 4-class HER2 testing results using the ensemble voting approach with the blue light illumination only. (C) ROC curves of the blind testing results using the ensemble voting approach with the blue light illumination only.

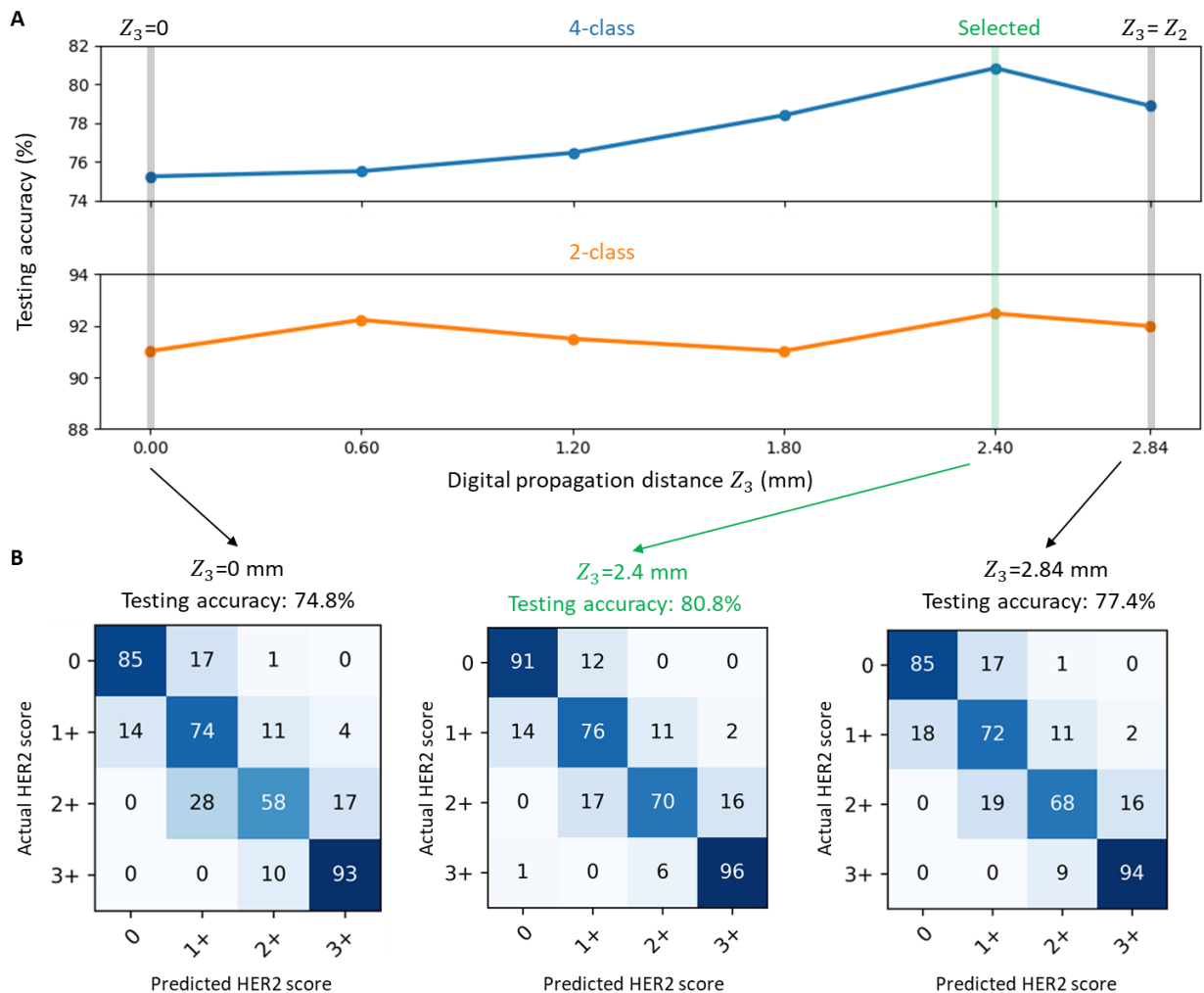

**Fig. 6. Impact of the digital propagation distance ($Z_3$) on the HER2 scoring accuracy.** (A) HER2 testing accuracy as a function of the digital propagation distance $Z_3$. (B) Confusion matrices of 4-class HER2 testing results using $Z_3$=0 mm, 2.4 mm and 2.84 mm.

**Table 1.** Cost estimate of our lensfree holography setup for automated HER2 scoring.

| Component | Unit price | Number | Manufacturer and part # | Total price |
|---|---|---|---|---|
| Basler CMOS image sensor | $584.00 | 1 | Basler ace acA3800-14um USB 3.0 Camera | $689.00 |
| Stepper motor with lead screw for x-dimension | $65.95 | 1 | Stepper Motor with 28cm Lead Screw: Bipolar, 200 Steps/Rev, 42×38mm, 2.8V, 1.7 A/Phase | $65.95 |
| Stepper motor with lead screw for y-dimension | $60.95 | 1 | Stepper Motor with 18cm Lead Screw: Bipolar, 200 Steps/Rev, 42×38mm, 2.8V, 1.7 A/Phase | $60.95 |
| Arduino Micro | $22.80 | 1 | Arduino Micro | $22.80 |
| Stepper motor driver | $19.95 | 2 | AMIS-30543 Stepper Motor Driver Carrier | $39.90 |
| Mechanical supporting parts | $50.00 | / | Custom | $50.00 |
| 3D printing material | $50.00 | / | Stratasys, Ltd, Objet30 V3 | $50.00 |
| **Total** | | | | **$978.60** |